\documentclass[superscriptaddress,aps,prl,showkeys,showpacs,twocolumn,10pt]{revtex4}
\usepackage[OT1,T1]{fontenc}
\usepackage{graphicx}
\usepackage{rotating}
\begin{document}

\title{ABC Resonance Structure in the Double-Pionic Fusion to $^4$He}
\date{\today}

\newcommand*{\IKPUU}{Division of Nuclear Physics, Department of Physics and 
 Astronomy, Uppsala University, Box 516, 75120 Uppsala, Sweden}
\newcommand*{\ASWarsN}{Department of Nuclear Reactions, National Centre for 
 Nuclear Research, ul.\ Hoza~69, 00-681, Warsaw, Poland}
\newcommand*{\IPJ}{Institute of Physics, Jagiellonian University, ul.\ 
 Reymonta~4, 30-059 Krak\'{o}w, Poland}
\newcommand*{\PITue}{Physikalisches Institut, Eberhard--Karls--Universit\"at 
 T\"ubingen, Auf der Morgenstelle~14, 72076 T\"ubingen, Germany}
\newcommand*{\KETue}{Kepler Center for Astro and Particle Physics, University
  of  
 T\"ubingen, Auf der Morgenstelle~14, 72076 T\"ubingen, Germany}
\newcommand*{\MS}{Institut f\"ur Kernphysik, Westf\"alische 
 Wilhelms--Universit\"at M\"unster, Wilhelm--Klemm--Str.~9, 48149 M\"unster, 
 Germany}
\newcommand*{\ASWarsH}{High Energy Physics Department, National Centre for 
 Nuclear Research, ul.\ Hoza~69, 00-681, Warsaw, Poland}
\newcommand*{\IITB}{Department of Physics, Indian Institute of Technology 
 Bombay, Powai, Mumbai--400076, Maharashtra, India}
\newcommand*{\HISKP}{Helmholtz--Institut f\"ur Strahlen-- und Kernphysik, 
 Rheinische Friedrich--Wilhelms--Universit\"at Bonn, Nu{\ss}allee~14--16, 
 53115 Bonn, Germany}
\newcommand*{\IKPJ}{Institut f\"ur Kernphysik, Forschungszentrum J\"ulich, 
 52425 J\"ulich, Germany}
\newcommand*{\JCHP}{J\"ulich Center for Hadron Physics, Forschungszentrum 
 J\"ulich, 52425 J\"ulich, Germany}
\newcommand*{\Bochum}{Institut f\"ur Experimentalphysik I, Ruhr--Universit\"t 
 Bochum, Universit\"atsstr.~150, 44780 Bochum, Germany}
\newcommand*{\ZELJ}{Zentralinstitut f\"ur Elektronik, Forschungszentrum 
 J\"ulich, 52425 J\"ulich, Germany}
\newcommand*{\Erl}{Physikalisches Institut, 
 Friedrich--Alexander--Universit\"at Erlangen--N\"urnberg, 
 Erwin--Rommel-Str.~1, 91058 Erlangen, Germany}
\newcommand*{\ITEP}{Institute for Theoretical and Experimental Physics, State 
 Scientific Center of the Russian Federation, Bolshaya Cheremushkinskaya~25, 
 117218 Moscow, Russia}
\newcommand*{\Giess}{II.\ Physikalisches Institut, 
 Justus--Liebig--Universit\"at Gie{\ss}en, Heinrich--Buff--Ring~16, 35392 
 Giessen, Germany}
\newcommand*{\HepGat}{High Energy Physics Division, Petersburg Nuclear Physics 
 Institute, Orlova Rosha~2, 188300 Gatchina, Russia}
\newcommand*{\Katow}{August Che{\l}kowski Institute of Physics, University of 
 Silesia, Uniwersytecka~4, 40-007, Katowice, Poland}
\newcommand*{\IFJ}{The Henryk Niewodnicza{\'n}ski Institute of Nuclear 
 Physics, Polish Academy of Sciences, 152~Radzikowskiego St, 31-342 
 Krak\'{o}w, Poland}
\newcommand*{\HiJINR}{Veksler and Baldin Laboratory of High Energiy Physics, 
 Joint Institute for Nuclear Physics, Joliot--Curie~6, 141980 Dubna, Russia}
\newcommand*{\IITI}{Department of Physics, Indian Institute of Technology 
 Indore, Khandwa Road, Indore--452017, Madhya Pradesh, India}
\newcommand*{\NuJINR}{Dzhelepov Laboratory of Nuclear Problems, Joint 
Institute for Nuclear Physics, Joliot--Curie~6, 141980 Dubna, Russia}
\newcommand*{\KEK}{High Energy Accelerator Research Organisation KEK, Tsukuba, 
 Ibaraki 305--0801, Japan}
\newcommand*{\IMPCAS}{Institute of Modern Physics, Chinese Academy of 
 Sciences, 509 Nanchang Rd., 730000 Lanzhou, China}
\newcommand*{\ASLodz}{Department of Cosmic Ray Physics, National Centre for
 Nuclear Research, ul.\ Uniwersytecka~5, 90--950 {\L}\'{o}d\'{z}, Poland}

\newcommand*{\Del}{Department of Physics $\&$ Astrophysics, University of 
 Delhi, Delhi--110007, India}
\newcommand*{\Wup}{Fachbereich Physik, Bergische Universit\"at Wuppertal, 
 Gau{\ss}str.~20, 42119 Wuppertal, Germany}
\newcommand*{\SU}{Department of Physics, Stockholm University, 
 Roslagstullsbacken~21, AlbaNova, 10691 Stockholm, Sweden}
\newcommand*{\Mainz}{Institut f\"ur Kernphysik, Johannes 
 Gutenberg--Universit\"at Mainz, Johann--Joachim--Becher Weg~45, 55128 Mainz, 
 Germany}
\newcommand*{\UCLA}{Department of Physics and Astronomy, University of 
 California, Los Angeles, California--90045, U.S.A.}
\newcommand*{\Bern}{Albert Einstein Center for Fundamental Physics, 
 Fachbereich Physik und Astronomie, Universit\"at Bern, Sidlerstr.~5, 3012 
 Bern, Switzerland}

\author{P.~Adlarson}    \affiliation{\IKPUU}
\author{W.~Augustyniak} \affiliation{\ASWarsN}
\author{W.~Bardan}      \affiliation{\IPJ}
\author{M.~Bashkanov}   \affiliation{\PITue}\affiliation{\KETue}
\author{T.~Bednarski}   \affiliation{\IPJ}
\author{F.S.~Bergmann}  \affiliation{\MS}
\author{M.~Ber{\l}owski}\affiliation{\ASWarsH}
\author{H.~Bhatt}       \affiliation{\IITB}
\author{K.--T.~Brinkmann}\affiliation{\HISKP}
\author{M.~B\"uscher}   \affiliation{\IKPJ}\affiliation{\JCHP}
% ...


\collaboration{WASA-at-COSY Collaboration}\noaffiliation






\begin{abstract}
Exclusive and kinematically complete high-statistics
measurements of the double pionic fusion reaction $dd \to ^4$He$\pi^0\pi^0$
have been performed in the energy range 0.8 - 1.4 GeV covering thus
the region of the ABC effect, which denotes a 
pronounced low-mass enhancement in the $\pi\pi$-invariant mass spectrum. The
experiments were carried out with the WASA detector setup
at COSY. Similar to the observation in the basic $pn \to d \pi^0\pi^0$
reaction, the data reveal a correlation between the ABC effect
and a resonance-like energy dependence in the total cross section. The
maximum occurs at m~=~2.37 GeV + 2$m_N$, {\it i.e.} at the same position as in
the basic reaction. The observed resonance width $\Gamma \approx$ 160
MeV can be understood from broadening due to  Fermi motion of the nucleons in
initial and final nuclei together with collision damping. Differential cross
sections are described equally well by the hypothesis of a $pn$ resonance
formation during the reaction process. 
\end{abstract}

\pacs{13.75.Cs, 14.20.Gk, 14.20.Pt}

\maketitle

The ABC effect denotes a pronounced low-mass enhancement in the
$\pi\pi$-invariant mass spectrum of double-pionic fusion reactions and is
named after Abashian, Booth and Crowe \cite{abc}, who observed this effect for
the first time in a single-arm magnetic spectrometer measurement of the $pd
\to ^3$HeX reaction.
Recent exclusive and kinematically complete measurements of the basic $pn \to d
\pi^0\pi^0$ reaction revealed the ABC effect to be strictly correlated with a
resonance-like structure at a mass of 2.37~GeV in the integral cross
section \cite{mb,MB}. The data are consistent with a
$I(J^P) = 0(3^+)$ assignment for this resonance structure \cite{MB,CW}.

Observation of pronounced ABC effects has also been reported from recent
exclusive measurements of fusion reactions to He isotopes: $pd
\to ^3$He$\pi^0\pi^0$ and $\pi^+\pi^-$\cite{he3} as well as $dd \to
^4$He$\pi^0\pi^0$ and $\pi^+\pi^-$ \cite{SK}. If the hypothesis of a $pn$
resonance causing the ABC effect is correct, then we should also observe a
resonance-like energy dependence of the total cross section in all these cases.
In such a scenario the reaction would be driven by an active $pn$ pair as in
the basic double-pionic fusion reaction, but with the difference that there
are now one or two spectator nucleons which merge with the active ones to the
final He isotope.




In order to investigate this issue in a comprehensive way we measured the
energy dependence of the 
{\it isoscalar} double-pionic fusion process $dd \to ^4$He$\pi^0\pi^0$ with
the WASA detector including deuterium pellet target \cite{wasa} at COSY (FZ
J\"ulich) using
deuteron beam energies in the range $T_d$~=~0.8~-~1.4~GeV. That way the full
energy range, where the ABC effect was observed in previous inclusive
single-arm magnetic spectrometer measurements \cite{ba,ban,wur}, was covered.

The emerging $^4$He particles were registered in the forward detector of WASA
and identified by the $\Delta$E-E technique. The photons from the $\pi^0$
decay were detected in the central detector \cite{CB}. 
Consequently four-momenta were measured for all emitted particles of an
event. Together with the condition that two pairs of the detected photons have
to fulfill the $\pi^0$ mass condition we have 6 overconstraints for the
kinematic fit of an event. 

Reaction particles have been detected over the full solid angle
with the exception of those $^4$He ejectiles (lab angles < 3$^\circ$), which
escaped in the beam-pipe. 

The absolute normalization of the data was obtained by a
relative normalization to the $dd \to ^3$He$n$ reaction measured
simultaneously with the same trigger. Our results for this reaction in turn
have been calibrated to the values of Ref. \cite{biz}. The error bars shown in
Fig.~1 (solid: statistical; dotted: systematic) are dominated by the
systematic uncertainties involved in this procedure \cite{AP}.

\begin{figure} 
\centering
\includegraphics[width=0.99\columnwidth]{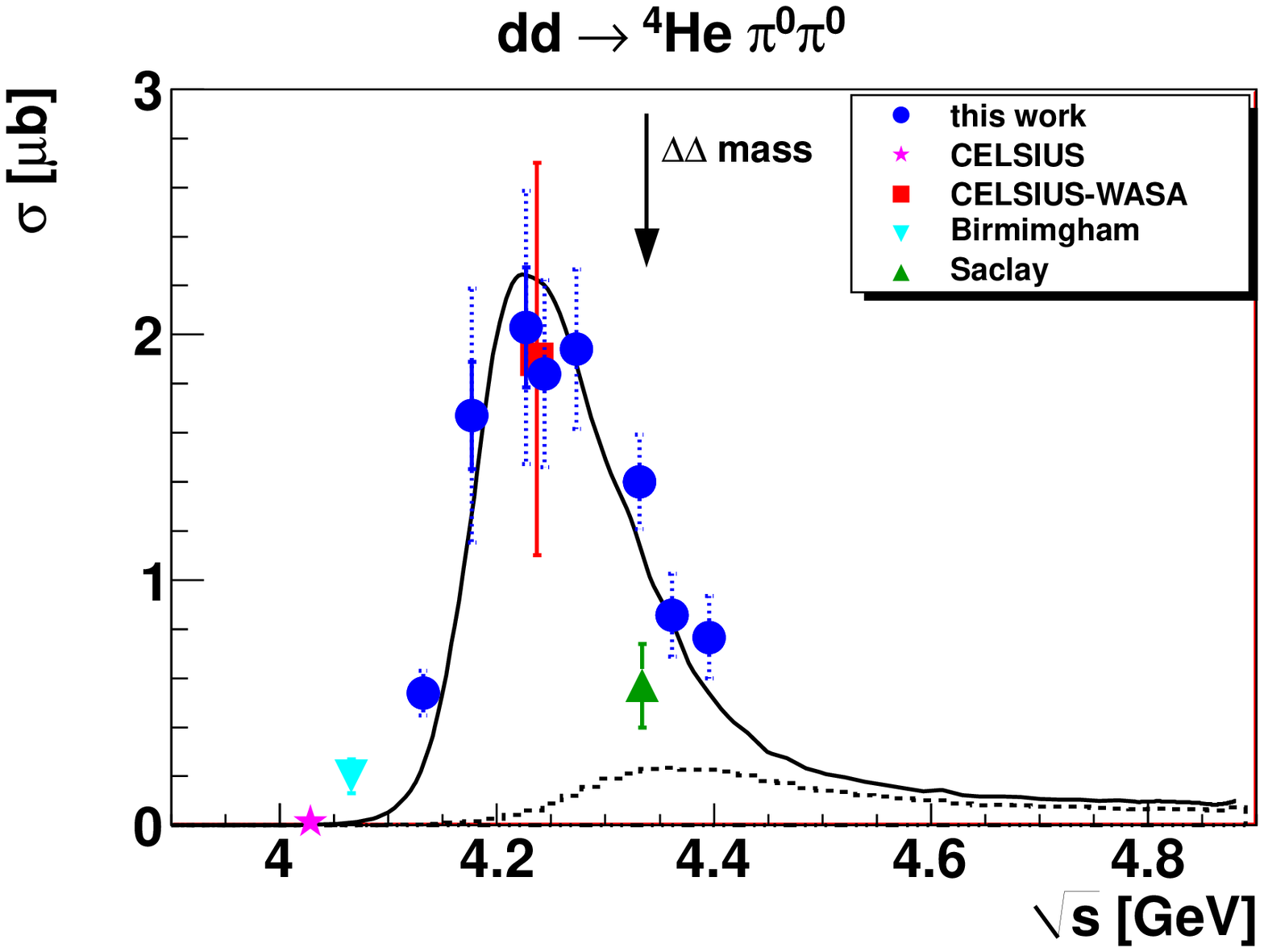}
\includegraphics[width=0.99\columnwidth]{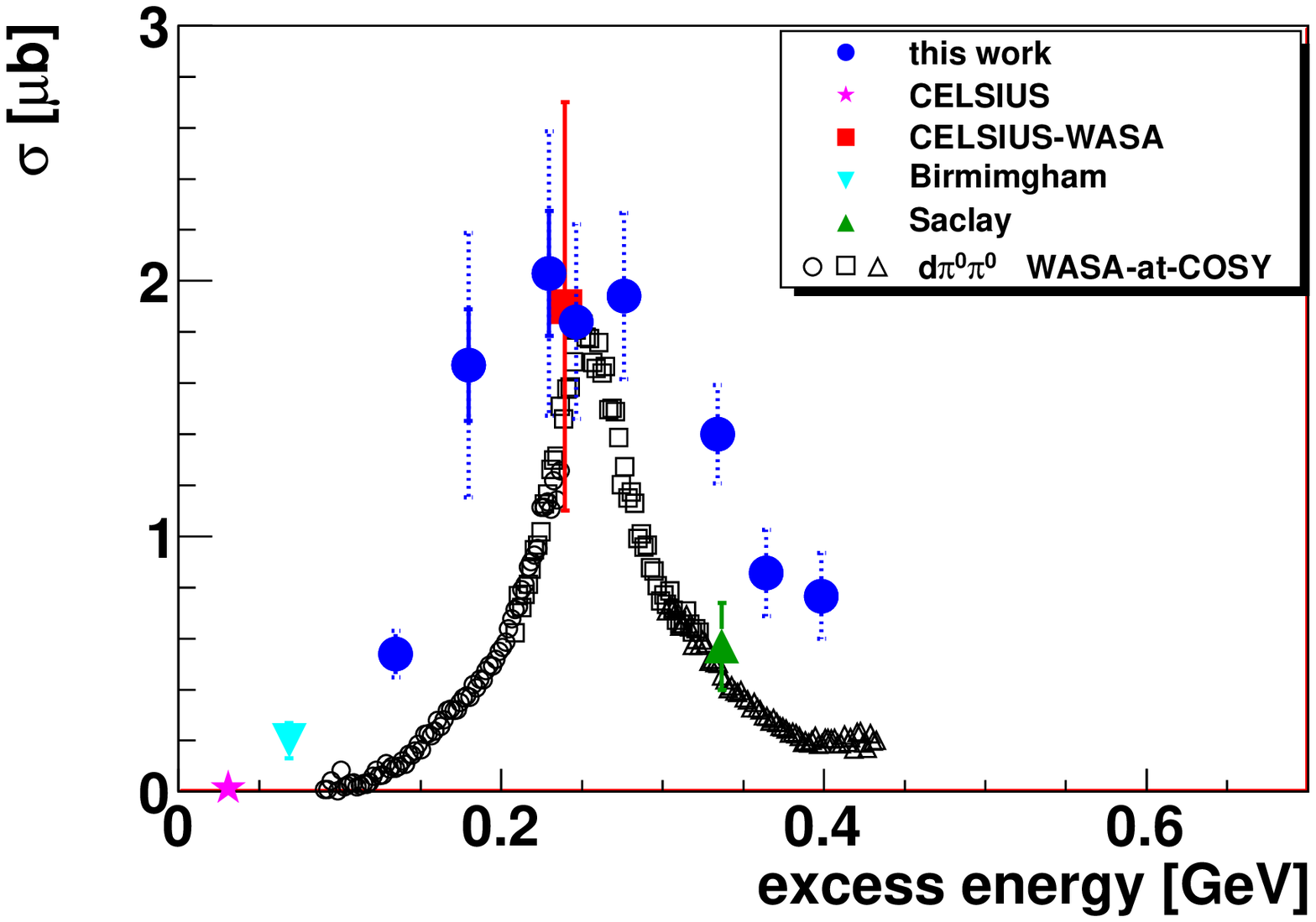}
\caption{\small {\bf Top:} Total cross section of the $dd \to ^4$He$\pi^0\pi^0$
  reaction in dependence of the center-of-mass energy $\sqrt s$. Results of
  this work (solid circles) are compared to results from a previous 
  exclusive measurement at CELSIUS/WASA \cite{SK} (square) as well as
  inclusive measurements at CELSIUS \cite{barg} (star), Birmingham \cite{cha}
  (inverted triangle) and Saclay \cite{ban} (triangle).  
  The drawn lines represent a conventional $t$-channel $\Delta\Delta$
  calculation \cite{MB} (dotted, scaled arbitrarily) as well as a
  calculation for an $s$-channel $pn$ resonance \cite{MB} with m = 2.37 GeV
  and $\Gamma_{pn}$~=~124~MeV (solid, 
  normalized to the data) including the Fermi motion of the
  nucleons in initial and final nuclei.
  {\bf Bottom:} Same as above, but now compared to the results for the basic
  $pn \to d \pi^0\pi^0$ (open symbols) reaction scaled down by a factor of 240.
}
\label{fig1}
\end{figure}

\begin{figure} 
\centering
\includegraphics[width=0.99\columnwidth]{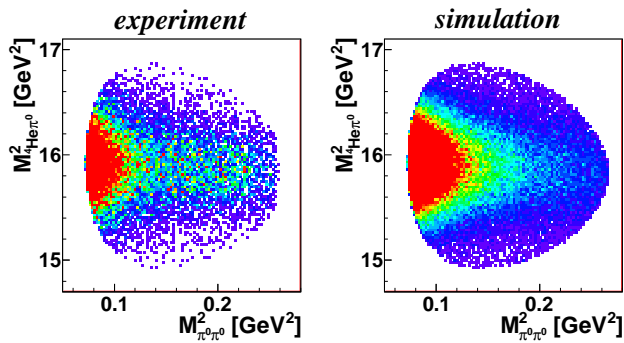}
\caption{\small
  Dalitz plot of $M_{^4He\pi^0}^2$ versus $M_{\pi^0\pi^0}^2$ at the energy of
  the peak cross
  section ($\sqrt{s}$ = 4.24 GeV. {\bf Left:} data ( highest
  $M_{\pi^0\pi^0}^2$ values cutted due to beam pipe, see text), {\bf right:} calculation
  for an $s$-channel $pn$ resonance with m = 2.37 GeV and $\Gamma$ = 124
  MeV including the Fermi motion of the nucleons in initial and final
  nuclei. 
}
\label{fig:spectra}
\end{figure}

\begin{figure} 
\centering
\includegraphics[width=0.49\columnwidth]{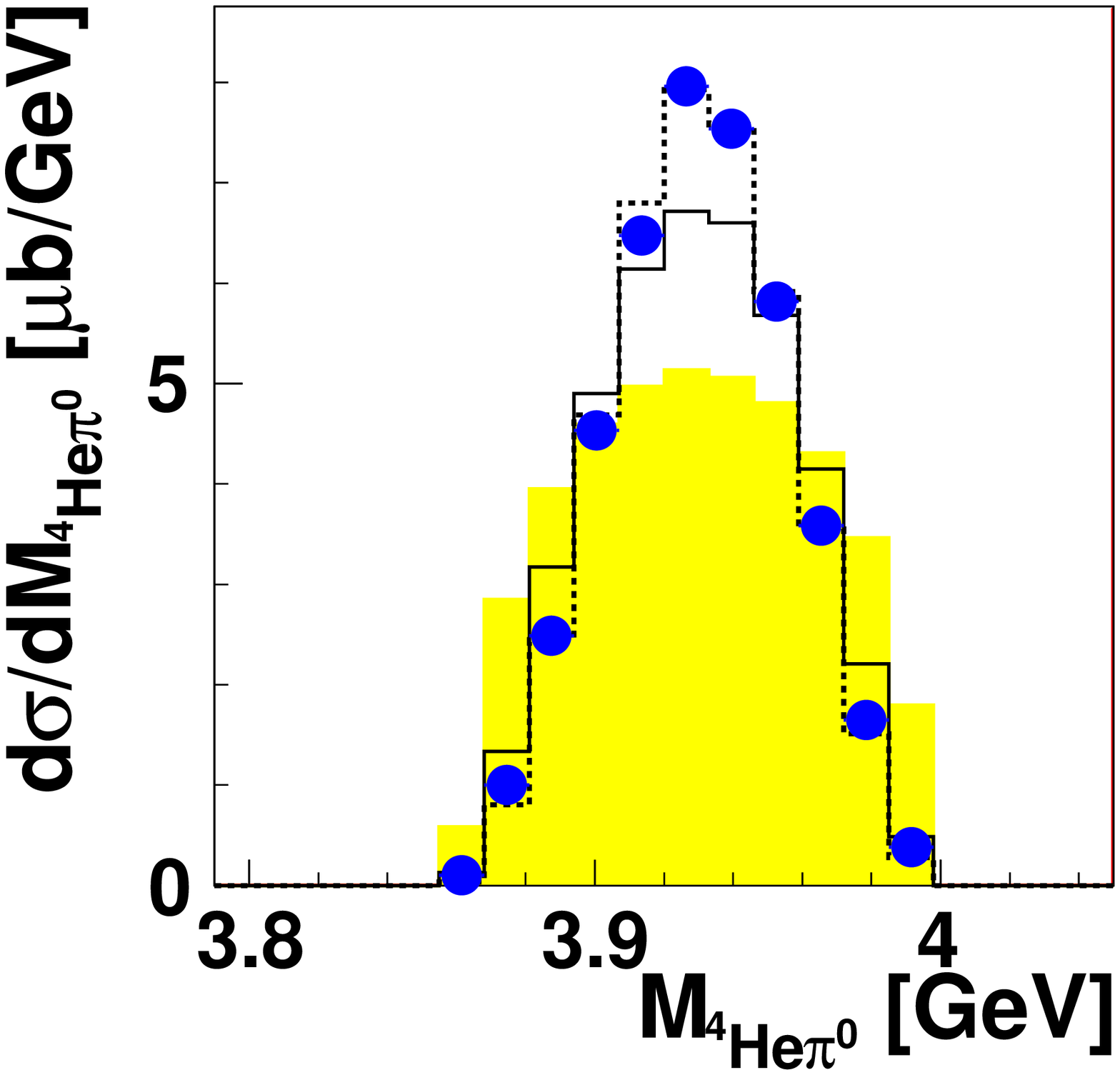}
\includegraphics[width=0.49\columnwidth]{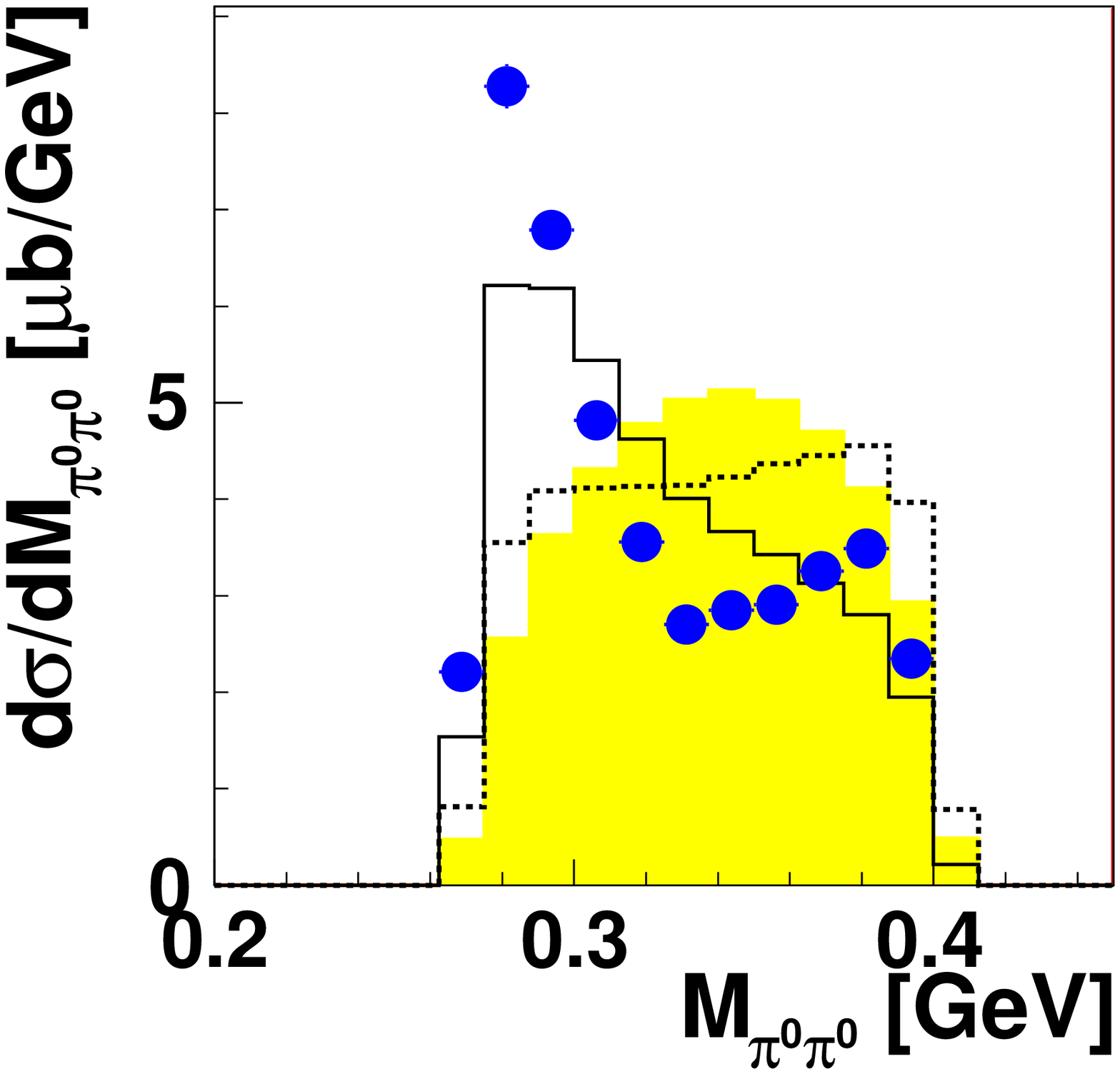}
\includegraphics[width=0.49\columnwidth]{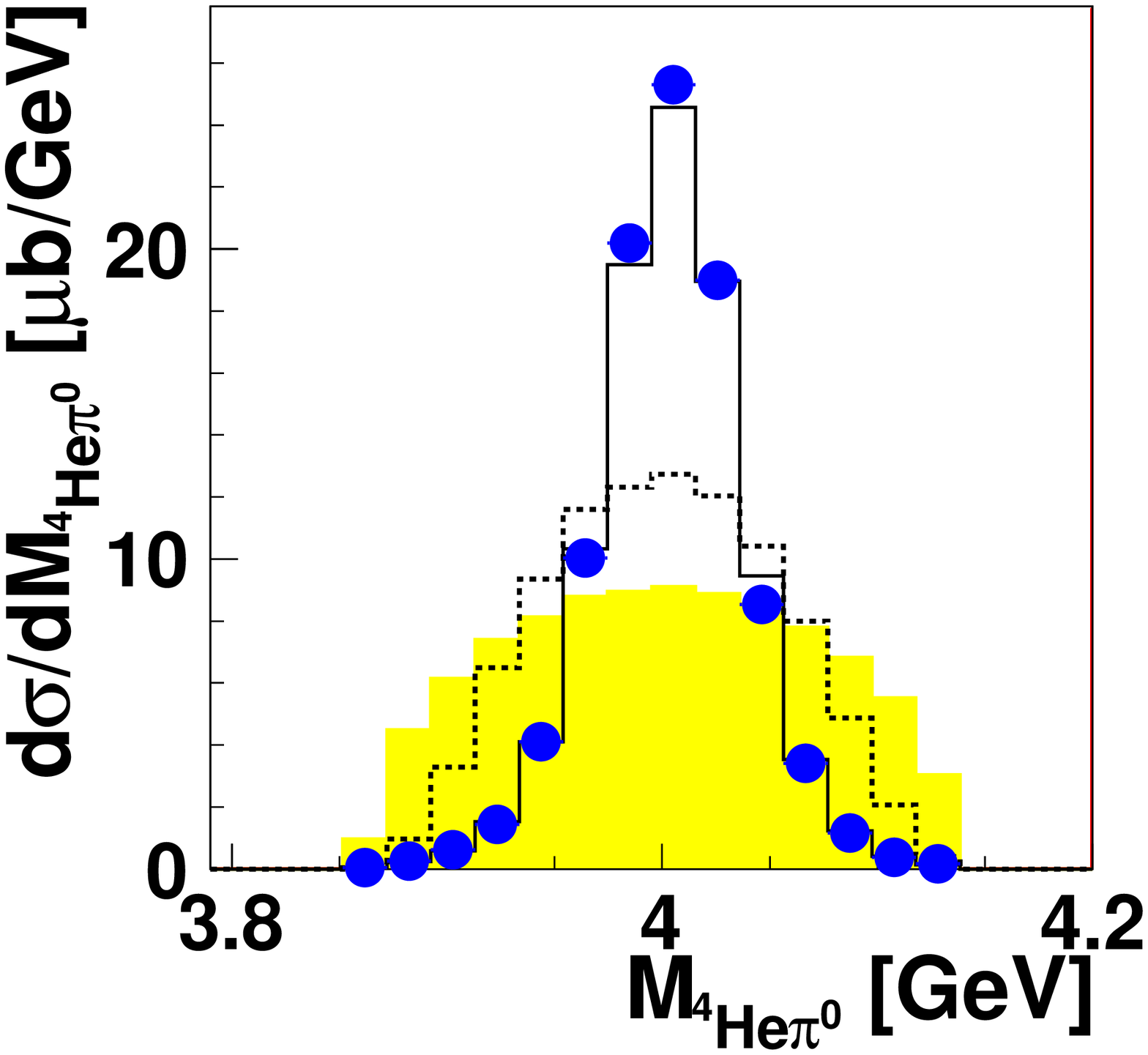}
\includegraphics[width=0.49\columnwidth]{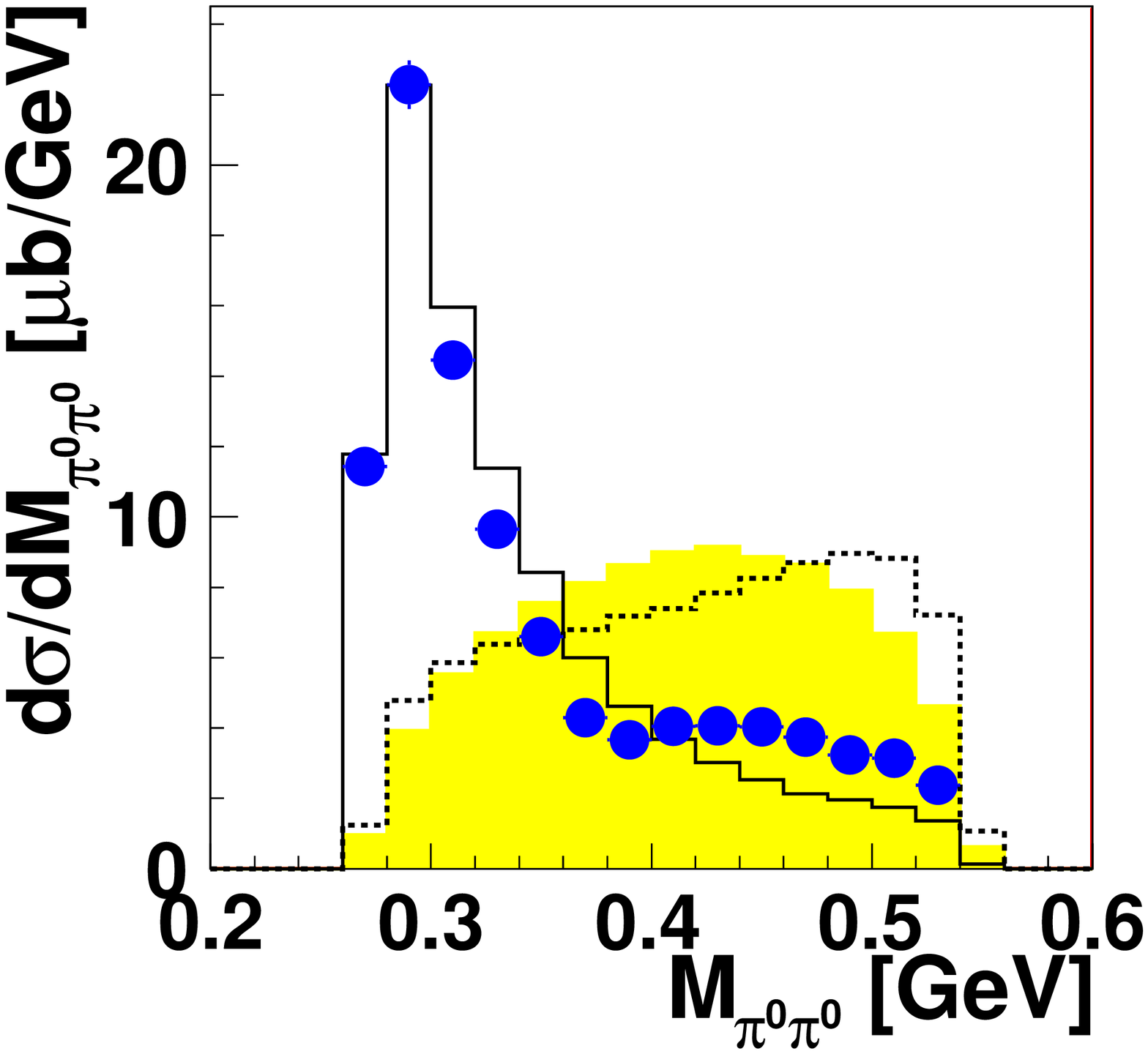}
\includegraphics[width=0.49\columnwidth]{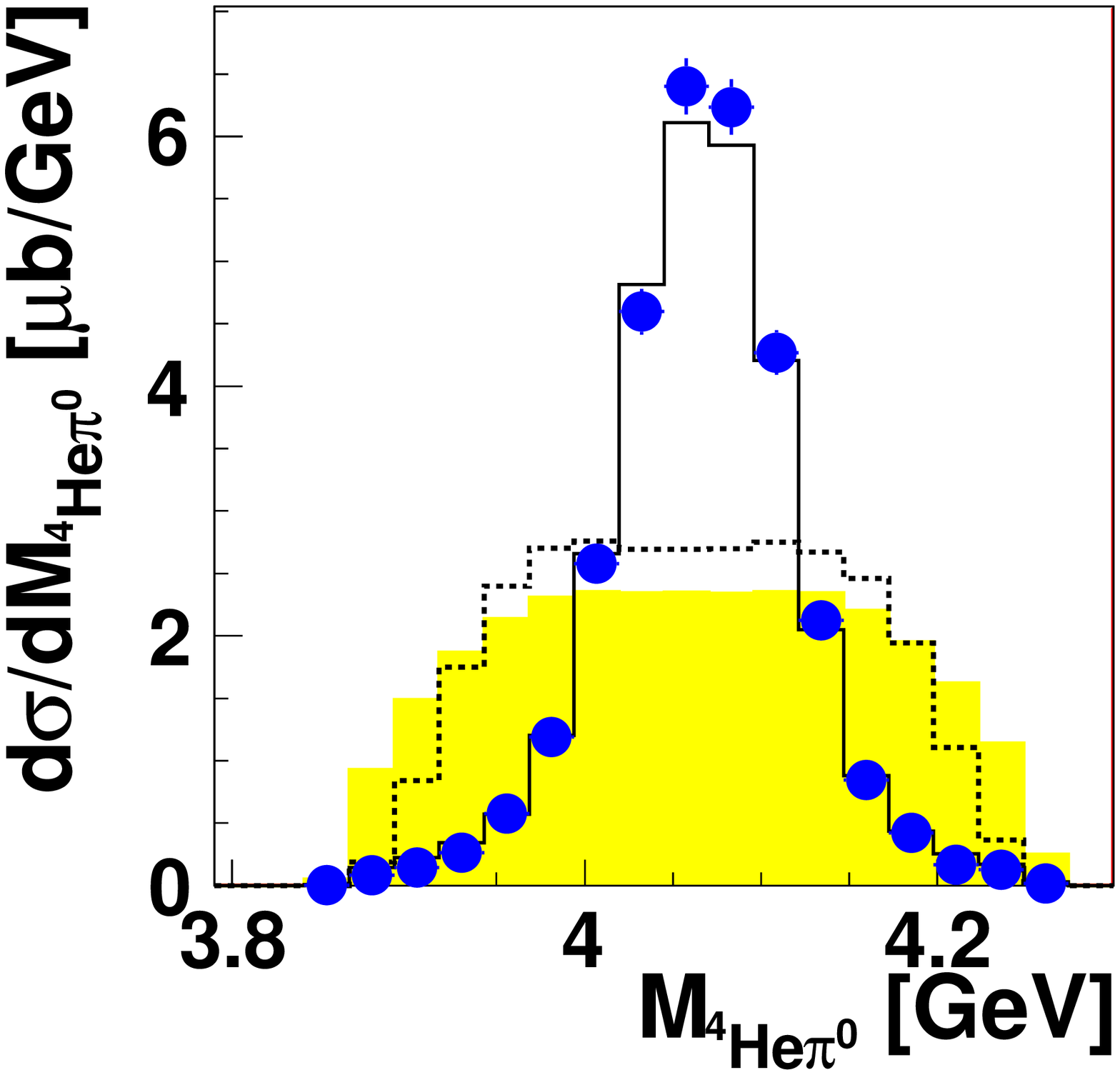}
\includegraphics[width=0.49\columnwidth]{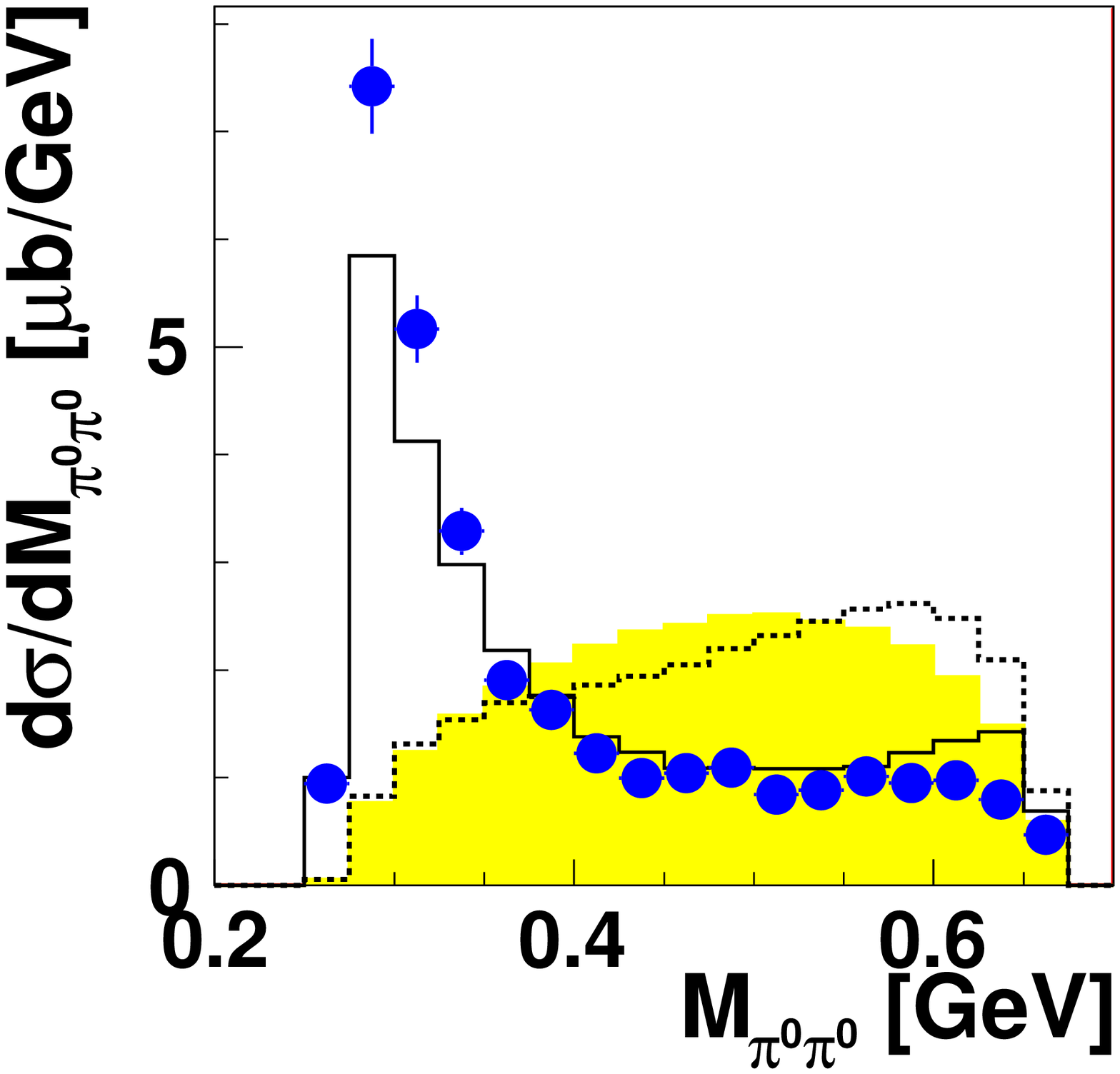}
\caption{\small 
  Distributions of $M_{^4He\pi^0}$ ({\bf left}) and $M_{\pi^0\pi^0}$ ({\bf right})
  at $\sqrt{s}$ = 4.13, 4.27 and 4.40 GeV. The shaded area denotes the
  phase-space distribution and the solid line a calculation of an $s$-channel
  $pn$ resonance with m = 2.37 GeV and $\Gamma$ = 124 MeV including the Fermi
  motion of the nucleons in initial and final nuclei. The dotted lines give a
  conventional $t$-channel $\Delta\Delta$ calculation.
}
\label{fig:spectra}
\end{figure}

The data obtained with WASA-at-COSY  are in reasonable agreement with the only
other exclusive measurement performed at CELSIUS at $T_d$ = 1.03 GeV
\cite{SK}. 

Results are shown in Figs.~1 - 4. 
Fig.~1 exhibits the measured energy dependence of the total cross section. In
the upper part the data from this and from previous work are plotted versus
the total 
energy $\sqrt s$ in the center-of-mass system (cms). With the exception of the
CELSIUS/WASA measurement \cite{SK} all other data originate from inclusive
single-arm magnetic spectrometer measurements \cite{barg,cha,ban}. The latter
include both the $\pi^0\pi^0$ and the $\pi^+\pi^-$ production channels. Since
both are purely isoscalar reactions, the cross sections of both channels
scale like 1:2, if we disregard the isospin violation due to different masses
of neutral and charged pions. Therefore the values from the inclusive
measurements \cite{barg,cha,ban} have been divided by three in Fig.~1. The data
point at $\sqrt s$ = 4.06 GeV is derived from a $0^\circ$-measurement assuming
isotropy. Since we measure strongly anisotropic $^4$He 
angular distributions even  at the lowest energies, we expect that actually
the total cross section is much lower than quoted in Ref. \cite{cha}.  

The total cross section data exhibit a
very pronounced resonance-like energy distribution. In the lower part of
Fig.~1 the data are plotted versus the excess energy of the reaction. That way
we can directly compare with the total cross section results for the basic $pn
\to 
d\pi^0\pi^0$ reaction. We see that the cross section maxima coincide in the
excess energy, though the width of the structure is much larger in the $^4$He
case. However, it is still substantially smaller than the width of about
$2\Gamma_\Delta$ of the conventional $t$-channel $\Delta\Delta$ process
(dotted lines in Fig. 1). This process peaks at about $2 m_{\Delta} + 2 m_N -
E_B$, where  $E_B$ denotes the nucleon binding energy difference between
$^4$He and the initial deuterons. As in the basic channel this occurs about 80
MeV above the observed maximum in the total cross section.

\begin{figure} 
\begin{center}

\includegraphics[width=0.69\columnwidth]{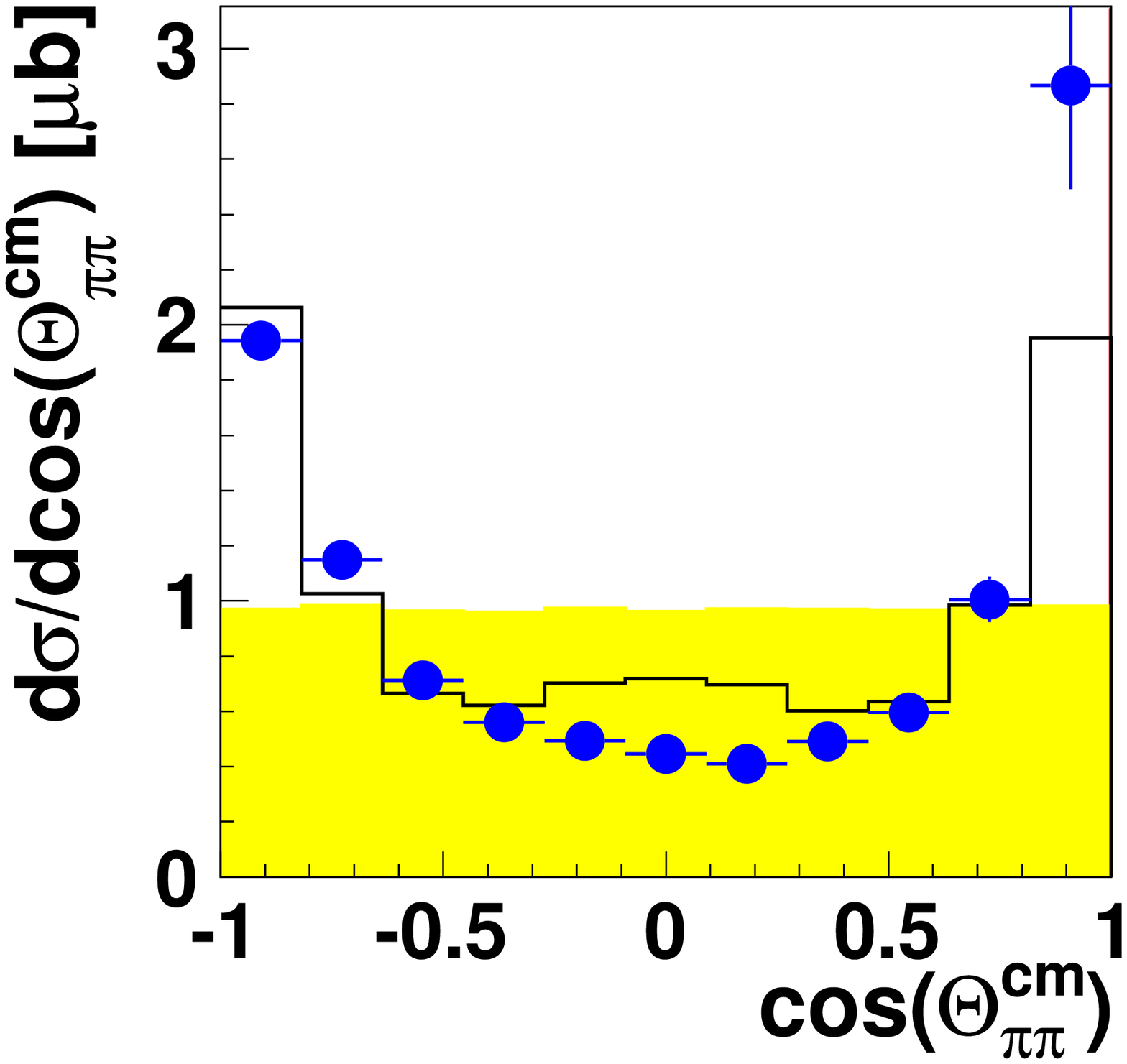}\\
\includegraphics[width=0.99\columnwidth]{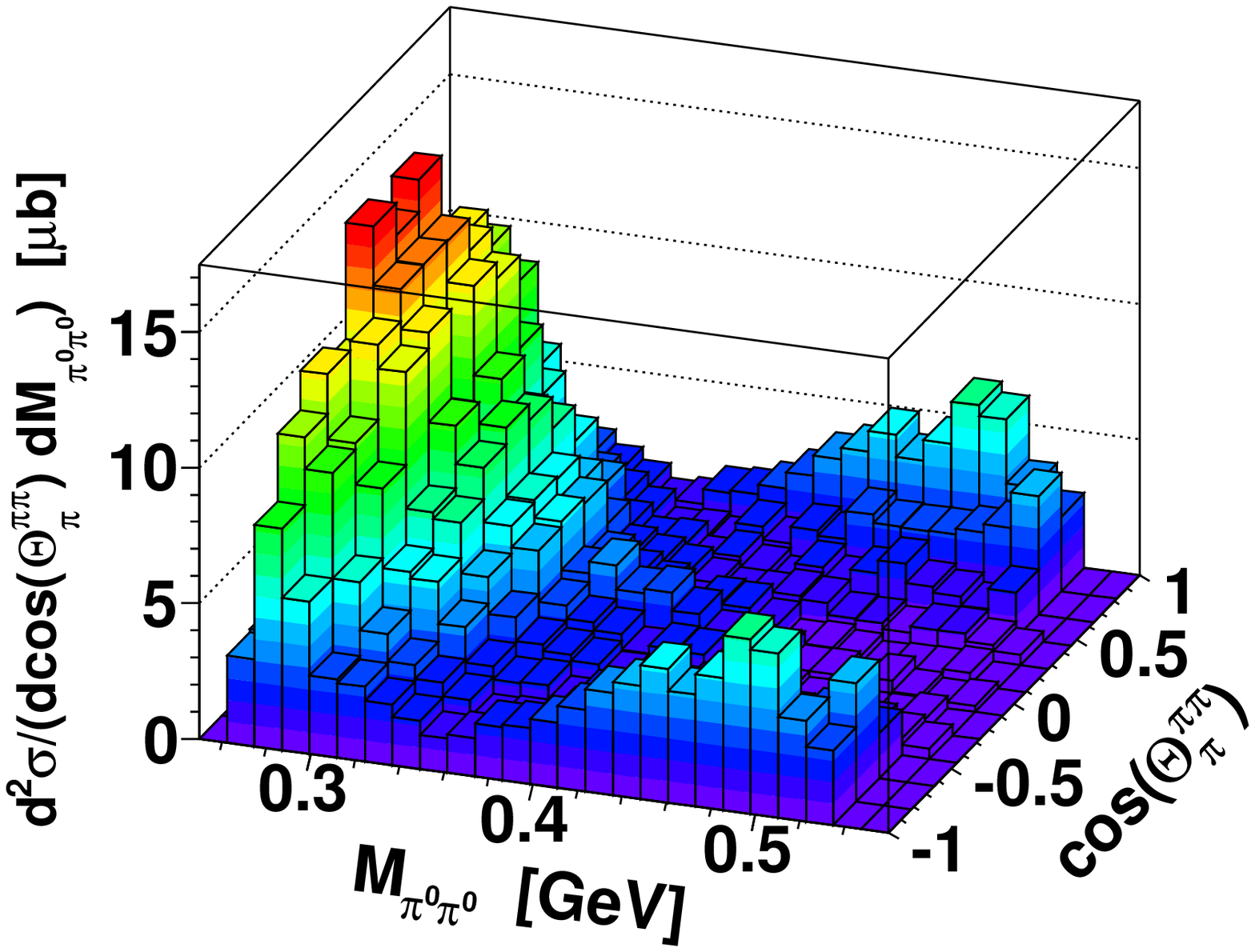}
\caption{\small Same as Fig. 3, but for angular distributions at the peak
  cross section ($\sqrt{s}$ = 4.27 GeV). 
  {\bf Top}: cms distribution of $\pi^0\pi^0$ system, {\bf bottom}:
  distribution of the pion angle $\Theta_{\pi^0}^{\pi^0\pi^0}$
  in the $\pi^0\pi^0$ subsystem (Jackson frame) in dependence of
  $M_{\pi^0\pi^0}$.
}
\end{center}
\end{figure}

We discuss in the following two invariant mass and two angular distributions,
which completely describe the 3-body reaction. 
The shape of all differential distributions remains rather unchanged over the
full energy region measured (cf. Fig.~3). 

Fig. 2 shows the Dalitz plots of the invariant mass
squared  $M_{^4He\pi^0}^2$ versus $M_{\pi^0\pi^0}^2$  
at the peak cross section ($\sqrt s$ = 4.24 GeV). The Dalitz plot is very
similar to that obtained in the basic reaction. It exhibits an
enhancement in horizontal direction, in the region of the $\Delta$
excitation, as it prominently shows up in the $M_{^4He\pi^0}$ spectra
displayed in
Fig. 3. This feature is consistent with the excitation of a
$\Delta\Delta$ system in the intermediate state -- as discussed for
the basic reaction \cite{MB}.
More prominent -- and even still much more pronounced than in the basic
reaction -- we observe here the ABC effect as a significant enhancement at the
low-mass kinematic limit of $M_{\pi^0\pi^0}$. Consequently the
Dalitz plot is mainly populated along the $\pi\pi$ low-mass border line. 
The $M_{\pi^0\pi^0}$ distribution is shown on the right side of Fig.~3 for
three different beam energies. It clearly exhibits the ABC effect at all
measured energies. We 
note, in passing, that we do not see a particular high-mass enhancement as the
inclusive data \cite{ban,ba,wur} suggest and as it was also predicted by model
calculations \cite{gar} specifically designed for the $^4$He case.
 
In Fig. 4 we show a selection of angular distributions. In the upper part the
angular distribution of the $\pi^0\pi^0$-system in the center-of-mass system
(cms) is shown. Since $\Theta_{\pi^0\pi^0} = 180^\circ - \Theta_{^4He}$ and
since the particles in the incident channel are identical, the cms
angular distribution must be symmetric about $90^\circ$ and the
$\pi^0\pi^0$-angular distribution is identical to the $^4$He-angular
distribution. The observed angular dependence is similar to the corresponding
one in the basic reaction, though significantly more peaked near $cos\Theta
= \pm 1$. At the bottom of Fig. 4 we show the lego plot of the distribution of
the pion angle $\Theta_{\pi^0}^{\pi^0\pi^0}$ in the $\pi^0\pi^0$-subsystem
(Jackson frame) in 
dependence of $M_{\pi^0\pi^0}$. This angular distribution is somewhat convex
curved in the ABC region (low-mass enhancement) as expected from the
$pn \to d\pi^0\pi^0$ reaction. After an intermediate
region, where the angular distribution flattens out, it gets very anisotropic
at high $M_{\pi^0\pi^0}$ values resembling a 
$d$-wave distribution. Since the cms $\pi^0$-angular distribution (see
Ref. \cite{SK}) exhibits a $p$-wave-like dependence as in the basic 
reaction \cite{MB}, the following picture emerges from the differential
distributions: In the ABC region the pions emerging from the $\Delta\Delta$
intermediate system couple to a $s$-wave pion pair, which is in relative
d-wave to the $^4$He system. At high $\pi^0\pi^0$-invariant masses the
situation is just reversed. In this picture the deuteron-like $np$ pair
resulting from the $\Delta\Delta$ decay merges with the passive $np$ pair to the
$^4$He ground state.

Since the features, which we observe here, are very similar to those observed
for the basic double-pionic fusion reaction, we adapt the ansatz used there
for the description of the $^4$He case \cite{AP}. There are only two
differences: First, 
the nucleons' momenta are smeared due to their Fermi motion in initial and
final nuclei. In particular the Fermi motion in the strongly bound $^4$He
nucleus leads to a substantial smearing of the energy dependence in the total
cross section adding nearly 40 MeV in the total width. To understand the full
observed width we need in addition a 
broadening of (124~-~68) MeV, which we ascribe to collision damping of the $pn$
resonance in the nuclear medium -- a feature well known, {\it e.g.}, from
$\Delta$ excitation in nuclei \cite{EW}.
Second, since the ABC effect appears even more concentrated
towards low $\pi^0\pi^0$ momenta in the
$^4$He case, we have to decrease the phenomenological cut-off parameter
\cite{MB} in the
$\Delta\Delta$ vertex function by about a factor of two. The resulting
calculation is shown in Figs.~1 - 4 by the solid lines providing a reasonable
description of the data. Since these calculations are semiclassical MC
simulations, the cut-off parameter is likely to fudge a number of shortcomings
in our simple treatment of the problem. Hence a full quantum mechanical
microscopic calculation would be very desirable.

In conclusion, our data on the double-pionic fusion to $^4$He establish the
correlation of a resonance-like energy dependence in the total cross section
with the ABC effect in very much the same way as shown for the basic
double-pionic fusion reaction to deuterium. A calculation based on the
$s$-channel $pn$ resonance with $I(J^P) = 0(3^+)$, m~=~2.37 GeV and $\Gamma$ =
124 MeV gives a good account of the observed distributions. The enlarged width
of the resonance-structure in the total cross section is explained by the
Fermi motion of the nucleons in initial and final nuclei as well as collision
damping. That way the ABC
effect in the double-pionic fusion to nuclei is traced back to a $pn$
resonance, which obviously is strong enough to survive even in the nuclear
medium.


We acknowledge valuable discussions with L. Alvarez-Ruso, A. Kudryavtsev, 
C. Hanhart, 
E. Oset,
A. Sibirtsev,  F. Wang and C. Wil\-kin on this issue. 
This work has been supported by BMBF, Forschungszentrum J\"ulich
(COSY-FFE), DFG, the Polish Ministry of Education, the Polish National
Science Centre,  
the Foundation for Polish Science (MPD) and EU (Regional Development Fund),
the Swedish Research Council and the Wallenberg foundation. 
We also acknowledge the support from 
the EC-Research Infrastructure Integrating Activity `Study of Strongly
Interacting Matter' (HadronPhysics2, Grant Agreement n. 227431) under the
Seventh Framework Programme of EU. 

\end{document}